\documentclass[english]{revtex4}
\usepackage{ae,aecompl}
\usepackage[T1]{fontenc}
\usepackage[latin9]{inputenc}
\usepackage{xcolor}
\usepackage{amsmath}
\usepackage{amssymb}
\usepackage{graphicx}
\PassOptionsToPackage{normalem}{ulem}
\usepackage{ulem}

\providecolor{lyxadded}{rgb}{0,0,1}
\providecolor{lyxdeleted}{rgb}{1,0,0}
\usepackage[mediumspace,mediumqspace,Grey,squaren]{SIunits}
\usepackage{ae}
\usepackage{aecompl}
\usepackage{url}
\usepackage{times}

\usepackage[a4paper]{geometry}

\usepackage{babel}
\begin{document}

\title{Graphene-Dielectric Composite Metamaterials: Evolution from Elliptic
to Hyperbolic Wavevector Dispersion and The Transverse Epsilon-Near-Zero
Condition}

\author{Mohamed A. K. Othman, Caner Guclu, and Filippo Capolino}

\email{fcapolino@uci.edu}

\affiliation{Department of Electrical Engineering and Computer Science, University
of California, Irvine, CA, $\mathit{92697}$, USA}
\begin{abstract}
We investigated a multilayer graphene-dielectric composite material,
comprising graphene sheets separated by subwavelength-thick dielectric
spacer, and found it to exhibit hyperbolic isofrequency wavevector
dispersion at far- and mid-infrared frequencies allowing propagation
of waves that would be otherwise evanescent in a dielectric. Electrostatic
biasing was considered for tunable and controllable transition from
hyperbolic to elliptic dispersion. We explored the validity and limitation
of the effective medium approximation (EMA) for modeling wave propagation
and cutoff of the propagating spatial spectrum due to the Brillouin
zone edge. We found that EMA is capable of predicting the transition
of the isofrequency dispersion diagram under certain conditions. The
graphene-based composite material allows propagation of backward waves
under the hyperbolic dispersion regime and of forward waves under
the elliptic regime. Transition from hyperbolic to elliptic dispersion
regimes is governed by the \textit{transverse} epsilon-near-zero
(TENZ) condition, which implies a flatter and wider propagating spectrum
with higher attenuation, when compared to the hyperbolic regime. We
also investigate the tunable transparency of the multilayer at that
condition in contrast to other materials exhibiting ENZ phenomena.
\end{abstract}

\keywords{Metamaterials, Multilayers, Plasmonics.}

\maketitle

\section{Introduction}

Hyperbolic metamaterial (HM) refers to a subcategory of uniaxially
anisotropic metamaterial, that can be modeled by a diagonal permittivity
tensor (in Cartesian coordinates) comprising entries with both positive
and negative real parts. The realization of hyperbolic dispersion
allows wave propagation over a wide spatial spectrum (infinite for
an ideal HM), that would be evanescent in a common isotropic dielectric
\cite{wavepropsmtith}. HMs are realized at optical frequencies using
metal-dielectric multilayers \cite{CanerPRBHM,Effectivemediumapproach,kidwaiHM},
or metallic nanowires \cite{Narimanovdarker}, and at terahertz and
infrared frequencies using semiconductor-dielectric multilayers \cite{Ciattoni,naik2010semiconductors}
or carbon nanotubes \cite{Nefedocindefinitecnt}. In multilayer HMs,
the emergence of hyperbolic dispersion does not rely on any resonant
feature, thus it poses a potential for broadband enhancement of the
local density of states (LDOS) \cite{jacob2010engineering}, subwavelength
imaging \cite{webbimaging,JacobnanophotonicsHM}, and lensing \cite{benedicto2012lens}.
Spontaneous emission rate of an emitter, as well as the radiative
decay of dye molecules, is proportional to the LDOS \cite{JacobnanophotonicsHM},
hence it can be substantially enhanced in the proximity of a hyperbolic
metamaterial \cite{kim2012improving,Krishnamoorthy13042012}. It was
demonstrated in \cite{CanerPRBHM} that the power scattered by a passive
nanosphere located in the proximity of a metal-dielectric HM is enhanced
by orders of magnitude, while the HM absorbs most of the scattered
power, opening a new frontier in super absorbers designs based on
near-fields transformation from evanescent to propagating regimes.
A wide band absorption was devised in \cite{Nefedovabsorption} using
tilted carbon nanotubes.

Multilayer HMs at optical frequencies take advantage of the wide frequency
band in which metals exhibit negative permittivity and support plasmonic
modes \cite{CanerPRBHM,Effectivemediumapproach}. At infrared frequencies,
graphene as a tunable inductive layer constitutes a potential building
block for multilayer HM realizations. Furthermore tunability of HMs
can be achieved using static fields to bias graphene \cite{Hansonmodesgraphene,vakil2011transformation}.
It is a remarkable material with a wide operational frequency band
starting from microwave regime \cite{Equivalentcctforgraphene}, through
terahertz frequencies \cite{tamagnone2012analysis}, and optical frequencies
\cite{ABajoabsorptiongeraphen}. Graphene was utilized in design of
metasurfaces in many different applications, such as polarizers and
absorbers \cite{aryatunable,FaradyrotJulien}, and cloaking devices
\cite{Aluenghetagraphenecloak}.

In this paper we investigate a graphene-dielectric multilayer material
that shows promising properties as tunable HM at far- and mid-infrared
frequencies, that was predicted to provide a large enhancement in
the Purcell factor \cite{BelovPRB,OthmanGrapheneHM}. In that recent
work, the enhancement of emitted power by electrically-small emitter
near the interface of graphene-based HM as well as the near-field
absorption properties were developed using effective medium approximation
(EMA) and transfer matrix methods, where the limitations and validity
of EMA were established \cite{OthmanGrapheneHM}. Here we show how
the wavevector dispersion diagram can be controlled and even transformed
between hyperbolic and elliptic curves at mid- and far-infrared regime.
Moreover, we demonstrate the design guidelines of the graphene-based
HM in terms of the physical parameters for the purpose of engineering
the evolution from hyperbolic to elliptic dispersion condition. .
In the last part of the paper we explore the \textit{transverse} epsilon
near zero (TENZ) condition, its relation to the dispersion diagram
and the enhanced transparency of a thin film made of TENZ graphene-dielectric
layers for TM waves with a wide range of incidence angle. The fabrication
of the metamaterial comprising as few as ten graphene-dielectric layers,
which were shown to have characteristics that resembles those of a
semi-infinite stack \cite{OthmanGrapheneHM}, could be realized utilizing
commercially available, high quality chemical-vapor-deposition-(CVD)-grown
graphene monolayer on a transition metallic (Ni or Cu) foil \cite{Largeareafewlayers,largeareaCu}.
from which graphene can be transferred onto a SiO$_{2}$/Si substrate
using an intermediate host such as a thermoplastic polymethyl-methacrylate
(PMMA) for enhancing the transfer process efficiency \cite{Largeareaconductiveelectrodes}.
This process is followed by depositing a thin film of SiO$_{2}$ or
SiC on the graphene flake using CVD. However, it was shown that a
graphene monolayer on SiO$_{2}$ can become highly disordered and
increases scattering losses \cite{dean2010boron}. The transfer of
few-layer graphene (FLG) \cite{Largeareafewlayers} on other compatible
materials such as Boron-Nitride (h-BN) might be of interest toward
realizing the metamaterial, since h-BN shares the same hexagonal structure
with graphene \cite{GrapheneBNhyperlens}.

\section{Effective medium analysis of graphene-dielectric multilayers}

\begin{figure}
\begin{centering}
\includegraphics[scale=0.25]{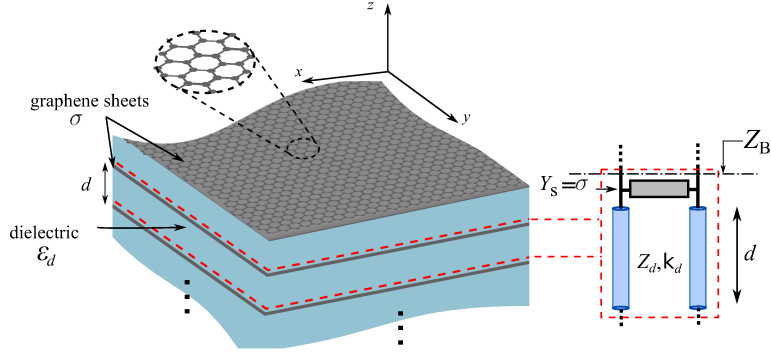} 
\par\end{centering}

\caption{\label{Fig__1}Graphene-dielectric multilayer HM topology, modeled
by a periodically-loaded transmission line. The unit cell is indicated
on the right and the graphene sheet is represented as a shunt admitta\textcolor{black}{nce,
and we denote the reference plane for evaluating the Bloch impedance.
At far- an}d mid-infrared frequencies, TM$^{z}$ waves exhibit hyperbolic
isofrequency wavevector dispersion.}
\end{figure}

Graphene is a one-atom-thick layer of hexagonal arrangement of carbon
atoms with a lattice constant of $0.264$ nm, hence spatial dispersion
effects introduced by graphene periodicity can be in general neglected
at terahertz frequencies. Although the existence of extremely slow
surface modes can trigger spatial dispersion effects \cite{Equivalentcctforgraphene,Meritmetadidlectric},
those modes are essentially highly evanescent due to the periodicity
of the multilayer structure studied here, as it will be shown in Sec. 3. Graphene is electrically modeled by the local isotropic sheet conductivity
$\sigma=\sigma'+j\sigma''$ (assuming time-harmonic variation of $e^{j\omega t}$),
which accounts for both interband and intraband contributions to the
total electronic transport \cite{electricfieldeffectnovoselov,sumrules}.
The sheet conductivity $\sigma$ is computed by the Kubo formula \cite{Hansondydicgraphetesigma},
which yields a function of frequency, chemical potential $\mu_{c}$,
phenomenological scattering rate $\Gamma$, and temperature $T$.
Here we assume for graphene $\Gamma=0.33$ meV (using the same notation
as in \cite{Hansondydicgraphetesigma}), which corresponds to a mean
electron scattering time of about $1$ ps, at room temperature $T=300$
K. Graphene supports relatively low loss TM plasmonic modes \cite{Hansonmodesgraphene}
(dictated by the negative imaginary part of the surface conductivity
$\sigma''<0$). As such, $\sigma''$, modeling the reactive response
of graphene, plays a fundamental role in the manifestation of hyperbolic
dispersion in multilayer graphene-dielectric materials, as described
in the following. We aim at analyzing an infinite periodic multilayer
structure depicted in Fig. \ref{Fig__1} whose unit cell is composed
of a graphene sheet and a dielectric layer of subwavelength thickness
$d$ and relative permittivity $\epsilon_{d}$. A physical understanding
of wave propagation in such multilayers with subwavelength period
can be established by using the effective medium approximation (EMA)
approach, which is a quasi-static or local approximation for metamaterials,
often adopted for metal-dielectric multilayers \cite{Effectivemediumapproach,CanerPRBHM,Meritmetadidlectric}.
According to EMA, the periodic multilayer is regarded as an anisotropic
homogeneous medium with effective relative permittivity tensor $\underline{{\pmb{\epsilon}}}_{\thinspace\textrm{eff}}=\epsilon_{t}(\hat{\mathbf{x}}\hat{\mathbf{x}}+\hat{\mathbf{y}}\hat{\mathbf{y}})+\epsilon_{z}\hat{\mathbf{z}}\hat{\mathbf{z}}$,
where the relative effective \textit{transverse} permittivity $\epsilon_{t}$
is found by averaging the transverse \textit{effective} displacement
current over the associated electric field in a unit cell (here, the
effective displacement current is defined as a quantity that includes
both displacement current in the dielectric slab and conduction current
in the infinitesimally-thin graphene sheet). Then the relative effective
permittivity parameter for transversely polarized field is 

\begin{equation}
\epsilon_{t}=\epsilon_{t}'-j\epsilon_{t}''=\epsilon_{d}-j{\displaystyle \frac{{\displaystyle \sigma}}{{\displaystyle {\displaystyle \omega}\epsilon_{0}d}}.}\label{Eq__1}
\end{equation}

Since an individual graphene sheet is infinitesimally-thin, the conduction
current is always along the sheet, hence the permittivity experienced
by $z-$directed electric field is not affected by graphene, leading
to $\epsilon_{z}=\epsilon_{d}.$ The relation in Eq. (\ref{Eq__1})
implies that when the graphene sheet is adequately inductive, in particular
when $\sigma''<-\omega\epsilon_{0}\epsilon_{d}d$, we obtain $\epsilon_{t}'<0$
and in turn the isofrequency wavevector dispersion is hyperbolic \cite{CanerPRBHM},
as demonstrated next. Let us consider plane waves propagating inside
the metamaterial with the spatial dependence $e^{-j\mathbf{k}.\mathbf{r}}$
where $\mathbf{k}=k_{x}\hat{\mathbf{x}}+k_{y}\hat{\mathbf{y}}+k_{z}\hat{\mathbf{z}}$
is the wavevector. A plane wave analysis is particularly useful in
understanding the multilayer's response to sources because the radiation
of a dipole inside or close to the metamaterial can be represented
as a spatial spectral sum of plane waves. Due to the symmetry of the
multilayer metamaterial with respect to the $z$ axis, we will use
$k_{t}=\sqrt{k_{x}^{2}+k_{y}^{2}}$ for denoting the transverse wavenumber
component and in the following $k_{t}$ is taken real representing
the spatial spectrum of TE$^{z}$ (electric field transverse to $z$)
and TM$^{z}$ (magnetic field transverse to $z$) waves. The $z$-directed
wavenumber $k_{z}=\beta_{z}-j\alpha_{z}$ can assume complex values
modeling propagation and attenuation, accounting also for natural
losses in the material constituents. Accordingly, the wavevector dispersion
of TE$^{z}$ and TM$^{z}$ waves inside the effective medium is given
as

\begin{eqnarray}
k_{z}^{2} & = & \epsilon_{t}k_{0}^{2}-k_{t}^{2},\qquad\textrm{TE}^{z}\label{Eq3_TEkz}\\
k_{z}^{2} & = & \epsilon_{t}k_{0}^{2}-\frac{\epsilon_{t}}{\epsilon_{d}}k_{t}^{2},\qquad\textrm{TM}^{z}\label{Eq4_TMkz}
\end{eqnarray}
 where $k_{0}=\omega\sqrt{\mu_{0}\epsilon_{0}}$ is the wavenumber
in free space. When the losses are neglected (i.e., if $\sigma'\rightarrow0$)
one would obtain purely real $\epsilon_{t}$, hence $k_{z}$ (obtained
via Eq.(\ref{Eq3_TEkz}) and Eq. (\ref{Eq4_TMkz})) assumes either purely
real values, denoting the propagating spectrum, or purely imaginary
values, denoting the evanescent spectrum. In this lossless case, hyperbolic
dispersion occurs when $\epsilon_{t}<0$, and the HM uniaxial medium
allows for propagation (i.e., $k_{z}$ is purely real) of extraordinary
waves (TM$^{z}$) with a large transverse wavenumber $k_{t}>\sqrt{\epsilon_{d}}k_{0}$;
these waves with $k_{t}>\sqrt{\epsilon_{d}}k_{0}$ would be otherwise
evanescent (i.e., $k_{z}$ is purely imaginary) either in a isotropic
dielectric with permittivity $\epsilon_{d}$, or in a generic uniaxial
anisotropic media with $\epsilon_{t}>0$. This unusual phenomenon
implies that high $k{}_{t}$ spectrum emanating from sources, which
would be evanescent in free space, can be converted to propagating
waves at HM interfaces. Ordinary waves (TE$^{z}$) are, however, evanescent
for any $k{}_{t}$ when $\epsilon_{t}<0$. On the other hand, when
$\epsilon_{t}>0$ we have real $k_{z}$ only for limited spectrum
of TM$^{z}$ waves with $k_{t}<\sqrt{\epsilon_{d}}k_{0}$, which leads
to the elliptic isofrequency wavevector dispersion. Therefore the
transition between hyperbolic to elliptic regimes is associated to
the condition $\epsilon_{t}=0$.

Instead, for realistic lossy cases, $k_{z}$ is complex and the wavevector
isofrequency dispersion becomes elliptic-like and hyperbolic-like
(for $\epsilon_{t}'>0$ and $\epsilon_{t}'<0,$ respectively), as
shown in the examples in next section. However, the interpretations
regarding propagation of power are still valid provided that losses
are relatively small, and we will show that moderate propagation losses
is a major advantage of graphene-based HMs at far- and mid-infrared
frequencies. When applying EMA, the dispersion relation $\beta_{z}-k_{t}$
is hyperbolic-like for $k_{t}>\sqrt{\epsilon_{d}}k_{0}$ when $\epsilon_{t}'<0$,
and it converges to the asymptote $\left|\beta_{z}\right|\approx\left|\epsilon_{t}'k_{t}/\epsilon_{d}\right|$
for large spatial wavenumber $k_{t}$, i.e., the $\beta_{z}-k_{t}$
dispersion becomes linear, with a slope of $\left|1+\sigma''/\left(\omega\epsilon_{0}\epsilon_{d}d\right)\right|$.

To validate our EMA hypothesis, we obtain a more accurate representation
of the wavevector dispersion relation by employing Bloch theory \cite{pozar2009microwave}
for a periodically loaded transmission line whose unit cell is illustrated
in Fig. \ref{Fig__1}. When each graphene sheet is modeled with a
complex admittance $Y_{s}=\sigma=\sigma'+j\sigma''$, the dispersion
relation for TM$^{z}$ or TE$^{z}$ waves in the periodic structure
is cast in the form

\begin{equation}
\cos k_{z}d=\cos\kappa_{d}d+j\frac{Y_{s}}{2}Z_{d}\sin\kappa_{d}d,\label{Eq_3_bloch-1}
\end{equation}
 where $\kappa_{d}=\sqrt{\epsilon_{d}k_{0}^{2}-k_{t}^{2}}$ is the
$z$-directed wavenumber of a wave inside the dielectric spacer, $Z_{d}^{\textrm{TM}}=\kappa_{d}/(\omega\epsilon_{0}\epsilon_{d})$
and $Z_{d}^{\textrm{TE}}=\omega\mu_{0}/\kappa_{d}$ are the characteristic
wave impedances for TM$^{z}$ and TE$^{z}$ waves, respectively. This
relation in (\ref{Eq_3_bloch-1}) is yet accurate for arbitrary $d$
and $k_{t}$, i.e., accounts for transverse wavenumber dispersion.
For the spectrum in which the dielectric layer's thickness is much
smaller than the Bloch wavelength and the wavelength inside the dielectric
itself ($\left|k_{z}d\right|\ll1,\left|\kappa_{d}d\right|\ll1$),
we can apply the following small argument approximations $\cos x\approx1-x^{2}$
and $\sin x\approx x$, the dispersion relation in Eq. (\ref{Eq_3_bloch-1})
simplifies to the one obtained via EMA in Eq. (\ref{Eq3_TEkz}) and Eq. (\ref{Eq4_TMkz})
using the same definitions for $\epsilon_{t}$ and $\epsilon_{z}$
\cite{OthmanGrapheneHM}. As we will discuss thoroughly in Sec. 3, Bloch theory proves that the propagating spectrum of TM$^{z}$
waves is limited due to the periodicity, manifested by the Brillouin
zone edge at which $\beta_{z}$=$\pm\pi/d$ , and therefore the propagating
spectrum in realistic HMs has an upper bound even in lossless cases.
Nevertheless, the Brillouin zone edge (i.e., $\beta_{z}$=$\pm\pi/d$
) is reached in general at higher values of $k_{t},$ provided that
the period $d$ is extremely subwavelength .

In the following we report some aspects that demonstrate the merits
of graphene-based HM: Graphene conductivity $\sigma=\sigma'+j\sigma''$
is tunable with chemical potential variation via electrostatic biasing,
hence $\epsilon_{t}'$ is also tunable through negative or positive
values, at a fixed frequency. This implies a possible transition between
hyperbolic to elliptic wavevector dispersion. The realization of HMs
using graphene is also prone to graphene's frequency response. For
instance, graphene sheets are mainly capacitive in mid- and near-infrared
frequencies, because intraband contributions in graphene are dominant,
and the $\textrm{TM}^{z}$ surface modes on a single graphene sheet
become on the improper Riemann sheet \cite{Hansonmodesgraphene}.
On the other hand, at very low frequencies (GHz regime), the interband
conductivity dominates leading to high losses. Hence a proper frequency
range for realizing hyperbolic dispersion extends from far-infrared
up to low mid-infrared frequencies. Furthermore, the dielectric thickness
also plays role on the frequency range of HM design. As the dielectric
thickness is increased, the frequency range of negative $\epsilon_{t}'$
shifts to lower frequencies which are undesirable due to significant
losses in graphene. Moreover, thicker spacers require a larger biasing
electrostatic potential between layers to achieve a moderate chemical
potential level in graphene sheets. On the other hand, when considering
smaller periods (in the range of several nanometers), it is expected
that graphene sheets are no longer electronically isolated for such
quantum-scale interspacing, and a tight binding model for graphene
layers must be taken into account in order to evaluate the conductivity
of graphene sheets \cite{Falkovseskymoonmulto,solidstatecommmultilayer}.
Therefore, for very small thicknesses, both EMA relation, reported
in (\ref{Eq4_TMkz}), and transfer matrix analysis must be modified
to account for quantum tunneling between graphene sheets. In the next
section we will explore and provide illustrative examples for graphene-based
HM designs in terms of frequency response, losses and tunability and
we will assess the validity of the EMA in predicting hyperbolic or
elliptic dispersion regimes.

\section{Hyperbolic and Elliptic Wavevector Dispersion}

\begin{figure}[b]
 \centering{}\includegraphics[scale=0.65]{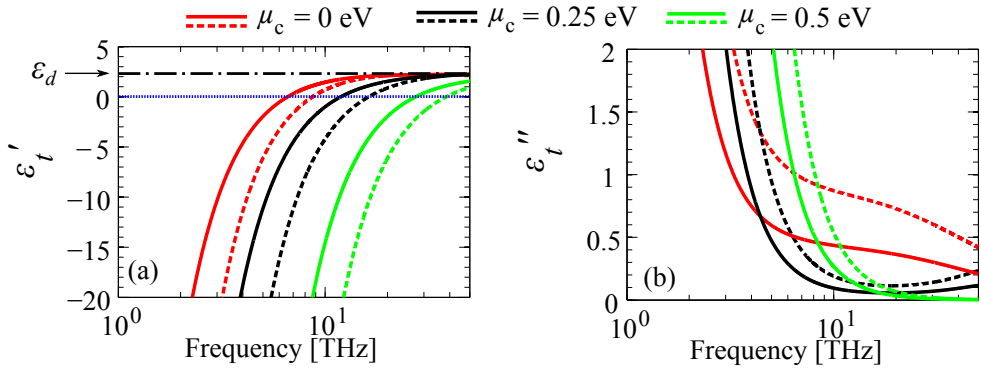}\caption{\label{Fig__2}Real and imaginary parts of the effective relative
\textit{transverse} permittivity $\epsilon_{t}=\epsilon_{t}'-j\epsilon_{t}''$
for graphene-based multilayer HM for two possible designs with $d=100$
nm (solid lines) and $d=50$ nm (dashed lines). }
\end{figure}

Let us consider a multilayer stack depicted in Fig. \ref{Fig__1},
that comprises graphene sheets and dielectric layers with $\epsilon_{d}=2.2$
and thickness $d.$ In our illustrations we only adopt positive values
for graphene chemical potential owing to the assumed reciprocity in
the multilayers, and consider a typical range for $\mu_{c}$ up to
0.5 eV in individual graphene sheets as suggested in \cite{Aluenghetagraphenecloak}.
We plot in Fig. \ref{Fig__2} the relative \textit{transverse} permittivity
$\epsilon_{t}=\epsilon_{t}'-j\epsilon_{t}''$ versus frequency, for
various chemical potential levels ($\mu_{c}=0$, 0.25, and 0.5 eV)
and dielectric thickness ($d=$100, 50 nm). First we observe that
the zero-crossing frequency of $\epsilon_{t}'$, where $\sigma''=-\omega\epsilon_{0}\epsilon_{d}d$,
is primarily defined by the period $d$ and it can be tuned via the
chemical potential; in turn the frequency of transition between the
hyperbolic and the elliptic dispersion regimes can be controlled.
Assuming $d=100$ nm (solid lines) in Fig. \ref{Fig__2}(a) we show
that the frequency at which $\epsilon_{t}'=0$ shifts from $6.6$
THz to $27.5$ THz by increasing the chemical potential from 0 eV
to 0.5 eV. For $d=$50 nm, similar control of the frequency at which
$\epsilon_{t}'=0$ is observed by varying $\mu_{c}$. Moreover when
$\mu_{c}=0$, we see that $\epsilon_{t}'=0$ occurs at 8.7 THz for
$d=$50 nm, a higher frequency than the $d=100$ nm case whose zero-crossing
frequency is around 6.6 THz. Graphene sheets become capacitive at
higher frequencies ($\sigma''=0$ denotes the transition from inductive
to capacitive, for instance, $\sigma''=0$ at $\simeq26$ THz when
$\mu_{c}=0$ eV), however its contribution to $\epsilon_{t}'$ becomes
negligible because of both $\omega$ in the denominator of (\ref{Eq__1})
and graphene conductivity saturates to $\pi e^{2}/(2h)\approx60$
$\micro$S with a very small imaginary part, and hence $\epsilon_{t}'$
approaches $\epsilon_{d}$. 

We show a relative variation in $\epsilon_{t}''$ when $\mu_{c}$
is increased, indicating a possible way to tune losses. Note that
when the frequency dependent \textit{transverse} permittivity $\epsilon_{t}'$
turns positive and becomes close to unity, satisfying $\sigma''\approx\omega\epsilon_{0}d(1-\epsilon_{d})$,
for instance at 15.6 THz when $\mu_{c}=$0.25 eV and $d=$100 nm,
a finite graphene-dielectric multilayer becomes almost transparent
to TE$^{z}$ and TM$^{z}$ plane waves in free space with $k_{t}\ll k_{0}$,
and all waves would travel with $k_{z}\approx k_{0}$, as seen from
(\ref{Eq4_TMkz}) when $\epsilon_{t}'\approx1$. 

\begin{figure}[tb]
 \centering{}\includegraphics[scale=0.6]{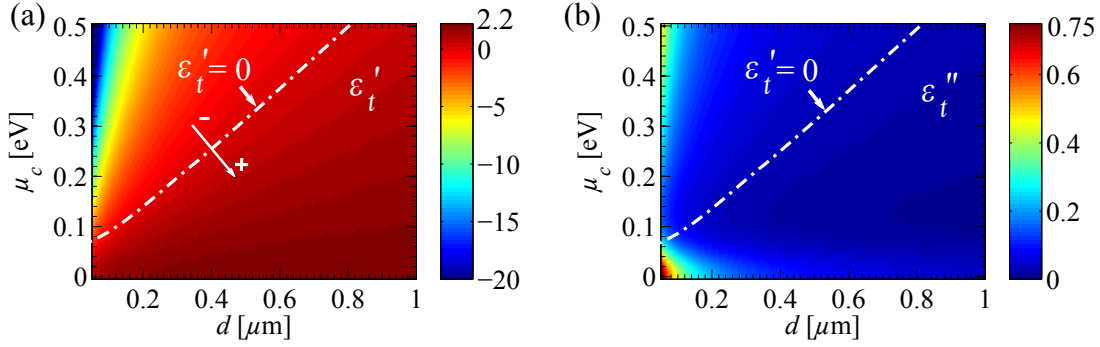}\caption{\label{Fig__3}Contour plot exploring the tuning capabilities of $\epsilon_{t}$
for graphene-based HM via chemical potential $\mu_{c}$ and dielectric
thickness $d$ at 10 THz.}
\end{figure}

In order to address some design considerations and tuning opportunities
of graphene-based HM, we show in Fig. \ref{Fig__3}(a,b), the real
and imaginary parts of $\epsilon_{t}$ as a colormap versus $\mu_{c}$
and $d$. We also indicate the $\epsilon_{t}'=0$ contour denoting
the transition between hyperbolic and elliptic dispersion regimes.
The selection of $d$ determines the range of chemical potential levels
in which hyperbolic/elliptic dispersion occurs. For instance, when
$d=0.2$ $\micro$m, a tuning range for hyperbolic dispersion starts
at $\mu_{c}=0.1$ eV, while for $d=0.6$ $\micro$m it begins at $\mu_{c}=0.35$
eV; this illustrates the need for thinner dielectric spacers due to
the limitations on the chemical potential levels' adjustability, up
to 0.5 eV in this paper. On the other hand, the choice of a thinner
dielectric spacer, i.e., smaller $d$, effectively induces higher
$\epsilon_{t}''$, so the losses embodied in $\epsilon_{t}''$ are
larger at the same frequency and bias. For example, when $d=0.1$
$\micro$m, $\epsilon_{t}''\simeq0.4$ but when $d=0.4$ $\micro$m
we notice that $\epsilon_{t}''\simeq0.2$, with larger negative $\epsilon_{t}'$
in the former case than in the latter. Nonetheless, a thin dielectric
spacer allows feasible biasing by standard values of static potential
\cite{aryatunable}. This demonstrates a basic trade-off in graphene-dielectric
HM design, between the tuning ranges, losses, and effective negative
values of $\epsilon_{t}'$, and leads to a broad interpretation of
the respective wavevector dispersion, as described next. 

\begin{figure}[t]
 \centering{}\includegraphics[scale=0.65]{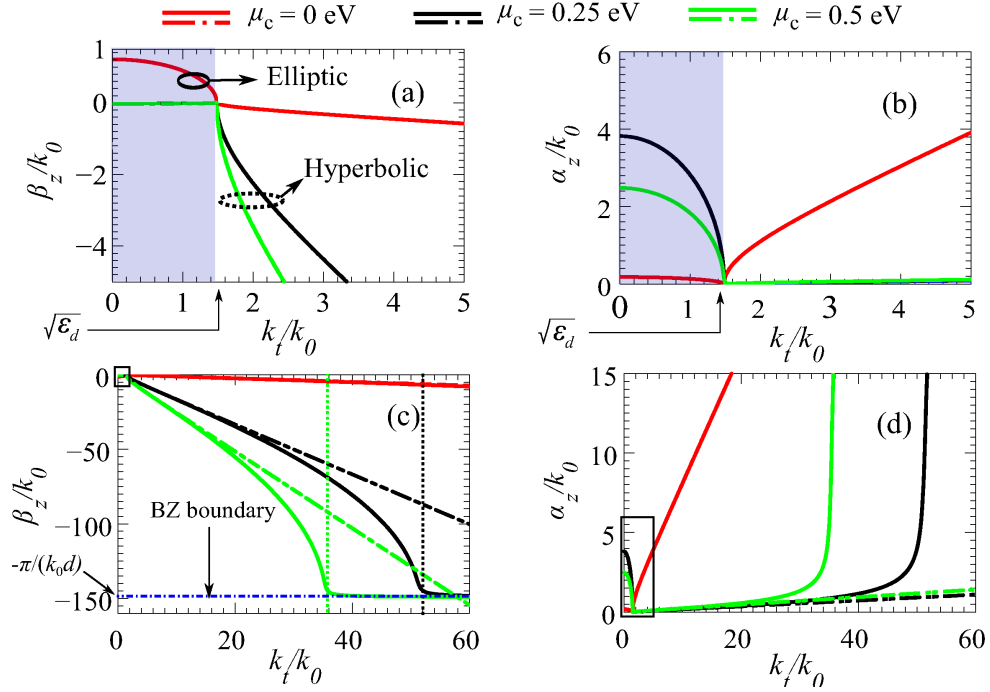}\caption{\label{Fig__4}Wavevector dispersion diagram of (a) $\beta_{z}$ and
(b) $\alpha_{z}$ versus $k_{t}$ (both normalized by $k_{0}$) at
10 THz and $d=100$ nm. In (c) and (d) a wider spatial spectrum of
the wavevector dispersion is provided in order to identify $k_{t}$
values where $\beta_{z}$ approaches the Brillouin zone edge ($\beta_{z}=-\pi/d$)
denoted by a horizontal dotted line in (c). This happens when $k_{t}\approx52k_{0}$
and $k_{t}\approx38k_{0}$ for $\mu_{c}=0.25$ eV and $\mu_{c}=0.5$
eV, respectively. Calculations are based on both EMA (dash-dotted
lines) and Bloch theory (solid lines).}
\end{figure}

The TM$^{z}$ wavevector dispersion diagrams according to EMA Eq. (\ref{Eq4_TMkz})
and Bloch theory for the multilayered medium Eq. (\ref{Eq_3_bloch-1})
are shown in Fig. \ref{Fig__4}. Here we report one of the two solutions
of Eq. (\ref{Eq4_TMkz}) and Eq. (\ref{Eq_3_bloch-1}) for $k_{z}=\beta_{z}-j\alpha_{z}$
that corresponds to a wave whose Poynting vector is directed towards
the $+z$ direction, noting that the other root $-k_{z}$ is also
a solution of (\ref{Eq4_TMkz}) and (\ref{Eq_3_bloch-1}), not reported
for symmetry reasons. Accordingly, the attenuation constant $\alpha_{z}$
has positive sign, associated to the field decay (due to possible
losses) along the +$z$ direction. On the other hand, for the hyperbolic
regime one observes $\beta_{z}<0$ indicating backward wave propagation
because it satisfies the backward wave condition $\beta_{z}\alpha_{z}<0$
explained in \cite{SalvoModesOpex2011}, for $k_{t}>\sqrt{\epsilon_{d}}k_{0}$.
 In general, for the elliptic case, when $k_{t}<\sqrt{\epsilon_{d}}k_{0}$
the valid $k_{z}=\beta_{z}-j\alpha_{z}$ solution with positive $\alpha_{z}$
is the one with $\beta_{z}>0$, indicating that waves under the elliptic
dispersion regime are forward waves because they satisfy the condition
$\beta_{z}\alpha_{z}>0$. In Fig. \ref{Fig__4}(c,d) we show the dispersion
diagrams in a much wider spatial spectrum than in Fig.\ref{Fig__4}(a,b)
for the same cases. In the reported cases, all with $d=100$ nm, $\beta_{z}$
curves in Fig. \ref{Fig__4} keep either an overall hyperbolic or
elliptic shape due to limited losses. When $\mu_{c}=$0 eV (and correspondingly
$\epsilon_{t}'>0$) the medium exhibits elliptic dispersion, moreover
$\beta_{z}$ is nonzero for $k_{t}>\sqrt{\epsilon_{d}}k_{0}$ where
$\alpha_{z}$ exhibits a dramatic increase, i.e., waves become mostly
evanescent. On the other hand, when $\mu_{c}=$0.25 eV or 0.5 eV,
one has $\epsilon_{t}'<0$ leading to hyperbolic dispersion. We emphasize
that EMA is fully capable of predicting the hyperbolic and elliptic
wavevector dispersion regimes in the spatial spectrum reported in Fig. 
\ref{Fig__4}(a,b) in perfect agreement with the Bloch wavenumber.
In a much wider range of the spatial spectrum $k_{t}$ as in Fig.
\ref{Fig__4}(c,d) the EMA-based normalized wavenumber $\beta_{z}/k_{0}$
starts to deviate from Bloch theory. Bloch theory predicts the band
edge where $\beta_{z}$ approaches $-\pi/d$ and $\alpha_{z}$ exhibits
a dramatic increase, denoting a bandgap. However EMA assumes infinite
growth of $\beta_{z}/k_{0}$ following the asymptotic linearized $\beta_{z}-k_{t}$
relation, given by $\beta_{z}\approx-\epsilon_{t}'k_{t}/\epsilon_{d}$
when $k_{t}\gg k_{0}$. For higher negative values of $\epsilon_{t}'$,
(corresponding to higher $\mu_{c}$), the Brillouin zone band edge
is met at smaller $k_{t}$ due to steeper $\beta_{z}-k_{t}$ curves,
as seen from Fig. \ref{Fig__4}(c) ($\epsilon_{t}'\simeq-1$ and $\epsilon_{t}'\simeq-11$
for $\mu_{c}=0.25$ and 0.5 eV). Although the effective permittivity
parameters are important for fast characterization of graphene-dielectric
composites and providing physical interpretation of the evolution
from elliptic to hyperbolic dispersion, they do not account for transverse
wavenumber dispersion \cite{chebykin2012nonlocal,Meritmetadidlectric}.
Accordingly, EMA predicts an indefinite propagating spatial spectrum
in HMs (that is indeed limited by Brillouin zone edge according to
Bloch model), and consequently overestimate the LDOS and the near-field
power absorption in HMs as already discussed in \cite{CanerPRBHM,Effectivemediumapproach,kidwaiHM,OthmanGrapheneHM}. 

\begin{figure}
\centering{}\includegraphics[scale=0.7]{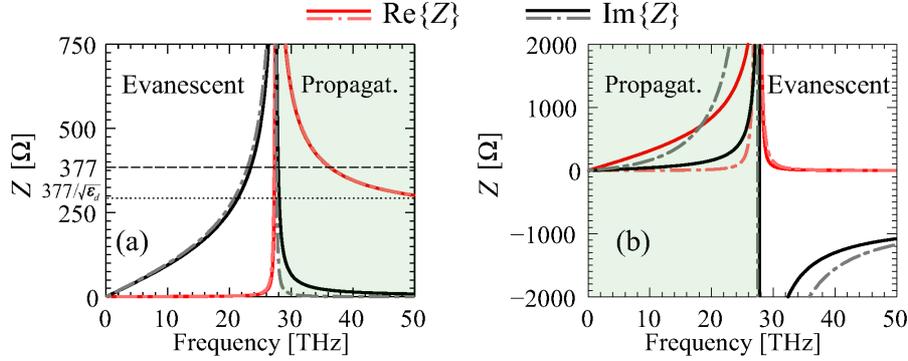}\caption{\label{Fig_5_Z}Real and imaginary parts of the Bloch (solid lines)
and effective (dashed lines) impedance of graphene-dielectric multilayers
with $d=100$ nm when $\mu_{c}=0.5$ eV for (a) $k_{t}=0$ and (b)
$k_{t}=5k_{0}.$ }
\end{figure}

\begin{figure}
\begin{centering}
\includegraphics[scale=0.7]{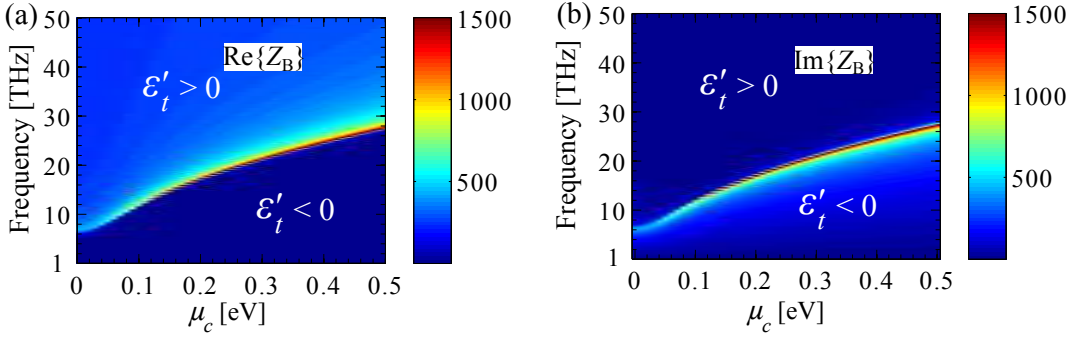}
\par\end{centering}

\caption{\label{Zbfreqmuc}(a) Real and (b) imaginary parts of the Bloch impedance
for $d=100$ nm and $k_{t}=0$.}
\end{figure}
We provide in Fig. \ref{Fig_5_Z} both the Bloch impedance of graphene-dielectric
multilayers at the reference plane shown in Fig. \ref{Fig__1}, with
$d=100$ nm. In addition, we report the effective wave impedance of
the metamaterial obtained via EMA for $\textrm{TM}^{z}$ plane wave,
$Z_{\textrm{eff}}=k_{z}/\left(\omega\epsilon_{0}\epsilon_{t}\right)$
where $k_{z}$ is evaluated using Eq. (\ref{Eq4_TMkz}), see \cite{FelsenMarcutiz}.
The two impedances are close to each other for $k_{t}=0$ case (Fig. \ref{Fig_5_Z}(a)) whereas for $k_{t}=5k_{0}$ the effective impedance
shows noticeable difference for both real and imaginary parts from
the Bloch calculations. Nonetheless, the effective impedance provides
a good prediction regarding the transition frequency between propagating
and evanescent spectra. Moreover, we notice that the real part of
the impedance is negligible at low frequencies in Fig. \ref{Fig_5_Z}(a),
whereas it peaks at the frequency where $\epsilon_{t}'=0$. From Fig.
\ref{Zbfreqmuc}(a) one can see that after $\epsilon_{t}'$ turns
positive, the impedance becomes dominantly real, with relatively small
reactive part, owing to the presence of a mainly propagating plane
wave in elliptic dispersion regime for $k_{t}=0$. On the contrary
for $k_{t}=5k_{0}$ case, at lower frequencies , wave propagates in
the hyperbolic dispersion regime while having $\epsilon_{t}'<0$,
and the impedance real part is relatively large, as depicted in Fig.
\ref{Fig_5_Z}(b), whereas the impedance becomes almost purely reactive
after $\epsilon_{t}'$ turns positive, denoting a mainly evanescent
wave. At higher frequencies, the impedance for $k_{t}=0$ case becomes
matched to free space at $\approx$37 THz at which $\epsilon_{t}'\approx1$
as shown in Fig. \ref{Zbfreqmuc}(a). At much higher frequency ranges,
the impedance approaches the impedance in isotropic lossless dielectric
where$\epsilon_{t}\approx\epsilon_{d}$ in both Fig. \ref{Fig_5_Z}(a)
and (b). For clarification, we report the Bloch impedance as a color
plot showing the dependance on frequency and chemical potential in
Fig. \ref{Zbfreqmuc}, where the impedance peaking is observed as
a clear manifestation of the TENZ condition, as it will be demonstrated
also in Sec. 4. Based on the conclusions in \cite{OthmanGrapheneHM},
in order to guarantee the validity of EMA for each spectral component
of propagating plane waves with $k_{t}<k_{0}$, the dielectric thickness
should be electrically-small, i.e., $d<0.02\lambda_{0}$ for accurate
representation of the impedance and wavevector using the homogenized
model derived above.

\begin{figure}[t]
 \centering{}\includegraphics[scale=0.57]{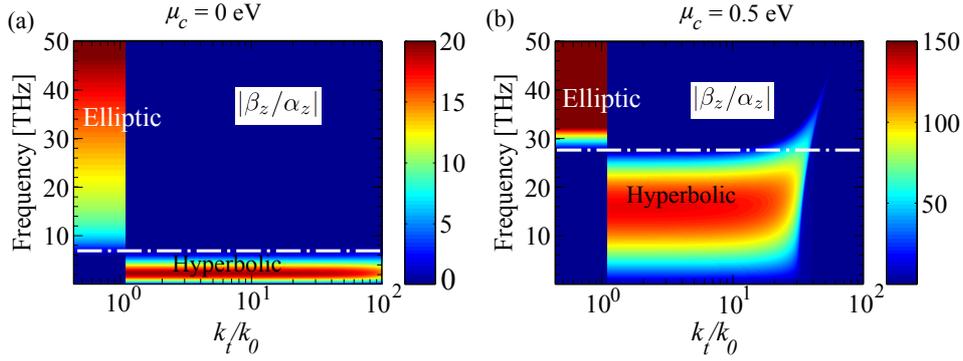}\caption{\label{Fig__5}The \textit{figure of merit} $\left|\beta_{z}/\alpha_{z}\right|$
versus frequency and spatial wavenumber $k_{t}$, for both hyperbolic
and elliptic regimes. Two chemical potential levels are considered:
(a) $\mu_{c}=$0 eV; and (b) $\mu_{c}=0.5$ eV.}
\end{figure}

We report in Fig. \ref{Fig__5} the frequency dependance of the quantity
$\left|\beta_{z}/\alpha_{z}\right|$ where $\alpha_{z}$ and $\beta_{z}$
are calculated by Bloch theory, for graphene-dielectric multilayers
with $d=100$ nm. The ratio $\left|\beta_{z}/\alpha_{z}\right|$ constitutes
a figure of merit for understanding if a wave is mainly propagating
or attenuating. The horizontal white dash-dotted line marks the transition
frequency from hyperbolic to elliptic dispersion (the latter occurring
always above the transition frequency) and the transition happens
when the real part $\epsilon_{t}'$ crosses zero and turns positive
causing the elliptic regime. For $k_{t}<\sqrt{\epsilon_{d}}k_{0}$,
$\beta_{z}$ is relatively very small compared to $\alpha_{z}$, which
implies mainly evanescent spectrum (purely evanescent in absence of
losses), for hyperbolic dispersion frequencies $\omega<-\sigma''/(\epsilon_{0}\epsilon_{d}d)$.
However, for $k_{t}>\sqrt{\epsilon_{d}}k_{0},$ wavevector dispersion
has a hyperbolic-like shape, with attenuation $\alpha_{z}$ moderately
low (and slightly increasing as seen in Fig. \ref{Fig__5}) due to
the losses in graphene, and therefore $\left|\beta_{z}/\alpha_{z}\right|$
exhibits an overall increase, where it reaches a maximum value $\simeq130$
as in $\mu_{c}=0$.5 eV yielding a wide propagating spectrum $\sqrt{\epsilon_{d}}<k_{t}/k_{0}\lesssim40$
at 10$-$20 THz. Notice that for even larger $k_{t}$, the propagation
constant $\beta_{z}$ tends to $-\pi/d$ while $\alpha_{z}$ experiences
an abrupt increase, as shown in Fig. \ref{Fig__4}(d), denoting the
beginning of a strong evanescent spectrum. In the elliptic dispersion
regime, 
occurring at higher frequencies such that $\omega>-\sigma''/(\epsilon_{0}\epsilon_{d}d)$,
the trend for $\beta_{z}$ and $\alpha_{z}$ is reversed. Elliptic
dispersion arises at 6.6 THz for $\mu_{c}=0$ eV, as depicted in Fig.
\ref{Fig__5}, and the propagating spectrum with $k_{t}<\sqrt{\epsilon_{d}}k_{0}$
is allowed in the composite multilayer. For higher chemical potentials,
as for example $\mu_{c}=0.5$ eV, hyperbolic wavevector dispersion
is supported for frequencies up to $27.4$ THz, and the dispersion
becomes elliptic thereafter. Notice that at frequencies less than
1 THz, waves poorly propagate due to higher losses in graphene sheets,
i.e., wave propagation has a low figure of merit. On the other hand,
elliptic dispersion regime, occurring for frequencies greater than
30 THz, has small attenuation constant for $k_{t}<\sqrt{\epsilon_{d}}k_{0}$
due to relatively low loss in graphene, and thus a high figure of
merit $\left|\beta_{z}/\alpha_{z}\right|>150$. Note that the lowest
operational frequency for hyperbolic dispersion regime with high $\left|\beta_{z}/\alpha_{z}\right|$
is limited by graphene losses, whereas the highest frequency is tunable
by the chemical potential. 
\begin{figure}[t]
 \centering{}\includegraphics[scale=0.57]{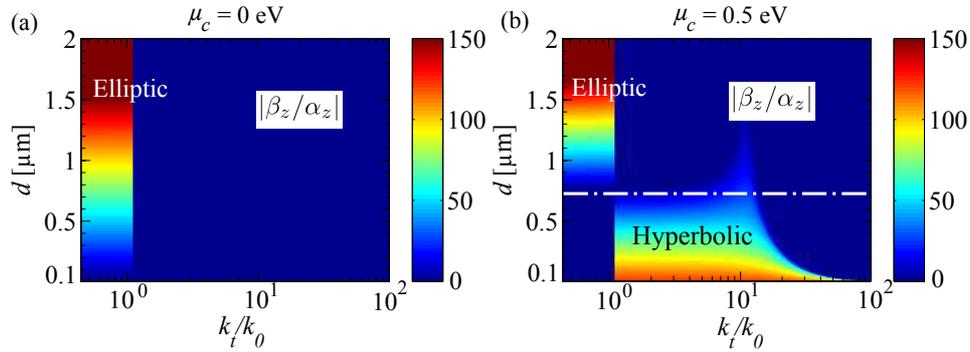}\caption{\label{Fig__6}The \textit{figure of merit} $\left|\beta_{z}/\alpha_{z}\right|$
versus dielectric thickness $d$ and spatial wavenumber $k_{t}$,
at 10 THz, for both hyperbolic and elliptic regimes. Two chemical
potential levels are considered: (a) $\mu_{c}=$0 eV; and (b) $\mu_{c}=0.5$
eV.}
\end{figure}

We now examine the how the figure of merit $\left|\beta_{z}/\alpha_{z}\right|$
varies versus the transverse wavenumber $k_{t}$, assuming different
design values for the dielectric spacing $d$. In Fig. \ref{Fig__6}(a)
we observe $\left|\beta_{z}/\alpha_{z}\right|$ at 10 THz varying
$d$, for $\mu_{c}=0$ eV, where only elliptic dispersion regime is
observed for any thickness $d$ considered. However, hyperbolic dispersion
is supported when appropriate chemical potential is achieved, as shown
in Fig. \ref{Fig__6}(b) for $\mu_{c}=0.5$ eV. In this latter case,
when $d=1$ $\micro$m, TM$^{z}$ waves are mainly evanescent for
large transverse wavenumber $k_{t}>\sqrt{\epsilon_{d}}k_{0}$, irrespective
of the chemical potential levels reported here. Consequently, a typical
dielectric thickness in the range of 50$-$100 nm is deemed appropriate
to utilize in graphene-dielectric multilayers for tunable HM designs.

\section{Transverse $\epsilon$-Near-Zero Condition}

\begin{figure}[b]
\centering{}\includegraphics[scale=0.7]{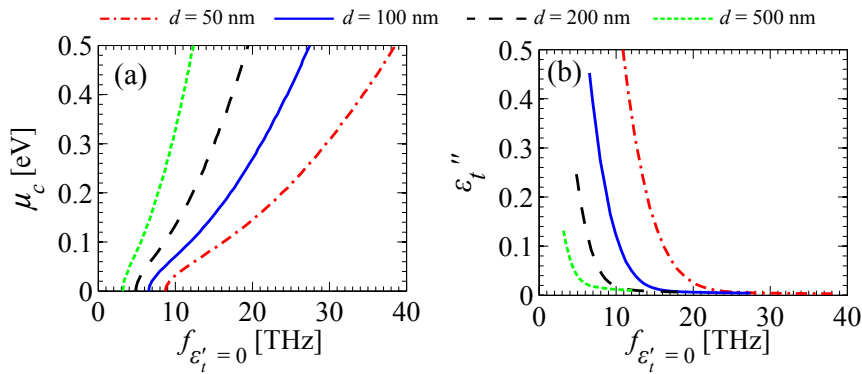}\caption{\label{fig:The-zero-crossing-frequency}The zero-crossing frequency
of $\epsilon_{t}'$ evaluated according to EMA $f_{\epsilon_{t}^{'}=0}=\sigma''/(2\pi\epsilon_{0}\epsilon_{d}d)$
varying the chemical potential, for various thicknesses $d$.
(b) Imaginary part of the transverse permittivity $\epsilon_{t}''$
evaluated at $f_{\epsilon_{t}^{'}=0}$. }
\end{figure}

Finally, we describe an interesting frequency region at which $\epsilon_{t}'$
changes sign and it assumes values very close to zero. We denote this
regime as transverse epsilon near zero (TENZ), which is manifested
under the condition $\sigma''\approx-\omega\epsilon_{0}\epsilon_{d}d,$
i.e., when graphene sheet's inductive susceptance compensates for
the small capacitive susceptance of each dielectric layer. We show
in Fig. \ref{fig:The-zero-crossing-frequency}(a) and (b), the level
of biasing potential ($\mu_{c}$) required to achieve the TENZ condition
at a given frequency,
and the corresponding $\epsilon_{t}''$, respectively. We note that
the required bias voltage for TENZ at a certain frequency decreases for thinner unit cells,
i.e., smaller $d$, however losses become larger due to increased
graphene sheet density, especially at low frequencies. For example
when $d=50$ nm, we require $\mu_{c}$ to be tuned to 0.1 eV in order
to achieve the TENZ condition at 15 THz, and we have $\epsilon_{t}''\approx0.1,$
whereas if the metamaterial is designed with $d=200$ nm, the amount
of bias required to realize TENZ condition at the same frequency is
about 0.2 eV and the losses are lower $\epsilon_{t}''\approx0.02$.
In view of such observations one can easily identify the tuning ranges
and show that for smaller unit cell thickness the tuning range is
larger but one must tolerate the losses in such design.

When considering wave propagation at that particular condition, and
if losses are to be neglected without compromising the generality
of the conclusions, the quasi-static approximation derived from EMA Eq. (\ref{Eq4_TMkz}) reveals a $\beta_{z}-k_{t}$ dispersion relation
with very small slope, i.e., the dispersion curve is almost flat.
However, at higher $k_{t}$ the EMA approximations become inaccurate,
and $\beta_{z}$ grows until it reaches the Brillouin zone edge $-\pi/d$.
The accurate wavevector dispersion of TM$^{z}$ waves according to
Bloch theory, using Eq. (\ref{Eq_3_bloch-1}) and $Z_{d}^{\textrm{TM}}=\kappa_{d}/(\omega\epsilon_{0}\epsilon_{d})$,
is given by 
\begin{eqnarray}
\cos k_{z}d & = & \cos\kappa_{d}d+j\frac{(\sigma'+j\sigma'')}{2}\frac{\kappa_{d}}{\omega\epsilon_{0}\epsilon_{d}}\sin\kappa_{d}d.\label{blochtenz-1}
\end{eqnarray}
The condition $\epsilon_{t}'\approx0$ is satisfied when $\omega\epsilon_{0}\epsilon_{d}d\approx-\sigma''$,
and it leads to

\begin{eqnarray}
\cos k_{z}d\approx\cos\kappa_{d}d+\frac{\kappa_{d}d}{2}\sin\kappa_{d}d+j\left|\frac{\sigma'}{2\sigma''}\right|\kappa_{d}d\sin\kappa_{d}d.\label{blochtenz}
\end{eqnarray}
This latter dispersion equation is further simplified under the small
argument approximation, $\left|\kappa_{d}d\right|\ll1$ as 
\begin{eqnarray}
\cos k_{z}d & \approx & 1+j\left(\kappa_{d}d\right)^{2}\left|\frac{\sigma'}{2\sigma''}\right|+O(\left|\kappa_{d}d\right|^{4}).\label{blochtenz2}
\end{eqnarray}

\begin{figure}[t]
\centering{}\includegraphics[scale=0.55]{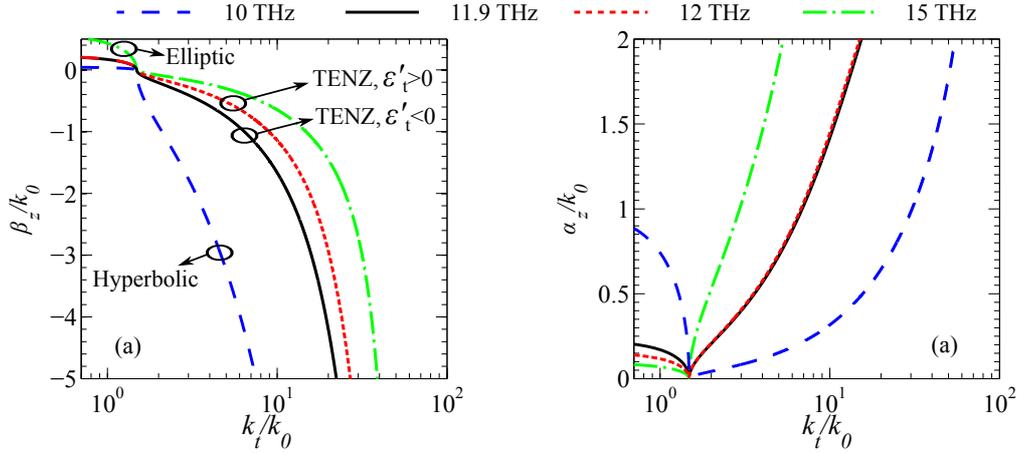}\caption{\label{Fig__7} Isofrequency wavevector dispersion in the TENZ, hyperbolic,
and elliptic regimes, showing both (a) $\beta_{z}$ and (b) $\alpha_{z}$
calculated by Bloch theory at four different frequencies (10, 11.9,
12, 15 THz), when $\mu_{c}=0.1$ eV.}
\end{figure}
The imaginary term in Eq. (\ref{blochtenz2}) is negligible since $\left|\left(\kappa_{d}d\right)^{2}\sigma'/2\sigma''\right|\ll1$
for graphene-dielectric multilayer with a subwavelength period, and
therefore one simply obtains $k_{z}\approx0$, far enough from the
Brillouin zone edge. Therefore, the TENZ condition $\epsilon_{t}'\approx0$,
implies a flat isofrequency dispersion diagram with small $k_{z}$
over a wide range of $k_{t}$. We report in Fig. \ref{Fig__7}(a,b)
the isofrequency wavevector dispersion at four different frequencies,
at which we show hyperbolic dispersion (10 THz with $\epsilon_{t}\simeq-1.01-j0.09$),
elliptic dispersion (15 THz with $\epsilon_{t}\simeq0.84-j0.05$),
and the TENZ transitional state (at 11.9 THz and 12 THz, with $\epsilon_{t}\simeq-0.001-j0.075$
and $\epsilon_{t}\simeq0.028-j0.072$, respectively), where both $\beta_{z}$
and $\alpha_{z}$ for all cases are normalized by $k_{0}$. In Fig.
\ref{Fig__7}(a) one can observe that the slope of the $\beta_{z}-k_{t}$
dispersion is reduced when $|\epsilon_{t}'|$ is much smaller than
unity, as also predicted analytically in Eq. (\ref{blochtenz2}), still
preserving limited values of the attenuation constant $\alpha_{z}$.
Note that the elliptic regime (at 15 THz) also shows a very low slope
of the $\beta_{z}-k_{t}$ dispersion, however the attenuation constant
$\alpha_{z}$ is large, because waves are mainly evanescent for large
$k_{t}$. Fig. \ref{Fig__7}(a) shows that the TENZ regimes are responsible
for almost flat propagation constant ($\left|\beta_{z}/k_{0}\right|<1$)
up to $k_{t}\simeq10k_{0}$, with a moderately low attenuation constant
$\alpha_{z}$. However, for larger $k_{t}$, we observe that $\beta_{z}$
experiences a sharp increase towards the Brillouin zone edge, together
with an increase of the attenuation constant $\alpha_{z}$. In Fig.
\ref{Fig__7}(b) we observe that the attenuation constant exhibits
significant difference for HM and TENZ regimes that requires some
important consideration. Although the two TENZ cases have smaller
$\epsilon_{t}''$ than the hyperbolic one (at 10 THz), they experience
a higher attenuation than HM case for $k_{t}>\sqrt{\epsilon_{d}}k_{0}$,
whereas the opposite relation is valid for $k_{t}<\sqrt{\epsilon_{d}}k_{0}.$
Therefore we can observe the two trends: on one hand TENZ allows flatter
$\beta_{z}-k_{t}$ relation and a wider $k_{t}$ spectrum than a fully
hyperbolic regime, on the other hand the hyperbolic regime exhibits
smaller attenuation constant $\alpha_{z}$ than the TENZ cases. Note
also that the TENZ is a transitional state toward elliptic dispersion,
at which the attenuation $\alpha_{z}$ becomes even higher for $k_{t}>\sqrt{\epsilon_{d}}k_{0}$,
and forward waves ($\beta_{z}\alpha_{z}>0$) can propagate for $k_{t}<\sqrt{\epsilon_{d}}k_{0}$
with low attenuation constant.

It has been shown in \cite{silveirinha2006tunneling,edwards2008experimental}
that isotropic epsilon-near-zero (IENZ) material inside a waveguide
supporting TE modes is able to tunnel electromagnetic waves. Here
we elaborate on TENZ materials at far- and mid-infrared frequencies
designed using graphene-dielectric multilayers and explore their capabilities
of tunneling electromagnetic waves \cite{LUOENZHM}. Consider an electrically-thin
slab of thickness $h$ made by either a TENZ ($\epsilon_{t}\approx0,$
$\epsilon_{z}\neq0$) or an IENZ ($\epsilon_{t}=\epsilon_{z}=\epsilon_{r}\approx0$)
material in free space. Under TE$^{z}$ wave incidence, TENZ and IENZ
slabs provide an identical response and the reflection from such slabs
can be set arbitrarily small by decreasing their thickness, as reported
in \cite{alu2007epsilon}. However, for $\textrm{TM}^{z}$ oblique
plane waves impinging on a lossless IENZ semi-infinite material, total
reflection occurs for angles greater than the critical angle $k_{t}^{c}/k_{0}=\sin\theta_{i}^{c}=\sqrt{\epsilon_{r}}\approx0$.
For an electrically-thin IENZ slab, transmission of TM$^{z}$ plane
wave takes place for small angles of incidence ($0$$<\theta_{i}<\theta_{i}^{c},$
where $\theta_{i}^{c}$ is considerably small) due to evanescent waves
exhibiting frustrated multiple reflections at the slab interfaces.
By including the effect of losses in IENZ slabs, absorption and local
electric field enhancement were reported for specific incident angles
$\theta_{i}>\theta_{i}^{c}$ in \cite{Salvoenz}. Instead, we provide
here the $\textrm{TM}^{z}$ reflection and transmission coefficients
($R_{\textrm{TM}}^{\textrm{TENZ}}$ and $T_{\textrm{TM}}^{\textrm{TENZ}}$)
for a thin TENZ slab

\begin{align}
R_{\textrm{TM}}^{\textrm{TENZ}} & =\frac{\zeta_{\textrm{}}^{\textrm{}}}{2Z_{0}+\zeta_{\textrm{}}^{\textrm{}}},\quad\quad T_{\textrm{TM}}^{\textrm{TENZ}}=\frac{2Z_{0}}{2Z_{0}+\zeta_{\textrm{}}^{\textrm{}}},\label{Eq_RTM}
\end{align}
where
\begin{equation}
\zeta_{\textrm{}}^{\textrm{}}=\frac{jh(k_{0}^{2}-k_{t}^{2}/\epsilon_{z})}{\omega\epsilon_{0}},\quad\quad Z_{0}=\frac{\sqrt{k_{0}^{2}-k_{t}^{2}}}{\omega\epsilon_{0}}.\label{eq:tenzpara}
\end{equation}
\begin{figure}[b]
\centering{}\includegraphics[scale=0.65]{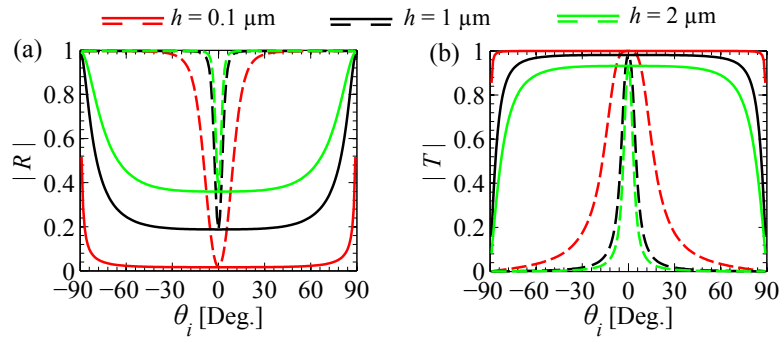}\caption{\label{fig:(a)-Reflection-and}Different characteristics of TM$^{z}$
plane wave (a) reflection and (b) transmission from a thin slab made
by a TENZ material (solid lines) and IENZ material (dashed lines)
at 37 THz. Material losses in this example are assumed negligible.
The TENZ material exhibits m\textcolor{black}{uch wider and flatte}r
parameters varying angle of incidence than the IENZ material.}
\end{figure}

\begin{figure}[t]
\begin{centering}
\includegraphics[scale=0.65]{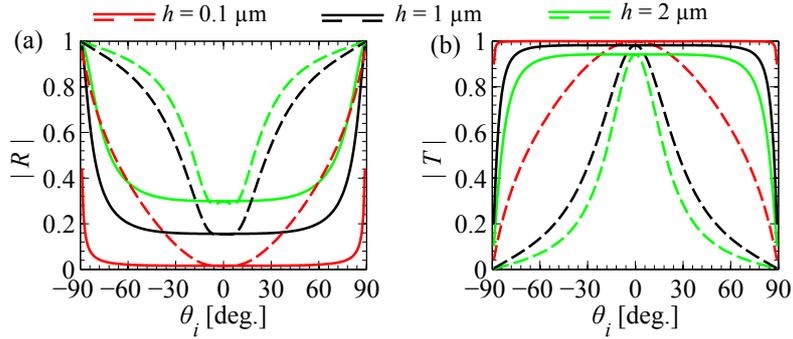}
\par\end{centering}

\caption{\label{fig:(a)-Reflection-and2}TM$^{z}$ plane wave (a) reflection
and (b) transmission from a slab made by graphene-dielectric layers
with $d=50$ nm and $h=Nd$ (solid lines, using transfer matrix analysis)
and an isotropic InAsSb slab of thickness $h$ (dashed lines) at 37
THz. }
\end{figure}

Therefore upon having a thin slab of TENZ material, $\zeta_{\textrm{}}$
can be made small enough (due to the existence of finite, non vanishing
$\epsilon_{z}$) in order to observe complete transmission for oblique
TM$^{z}$ waves with a wide range of incidence angles. This is in
contrast to what happens for the IENZ case with $\epsilon_{z}$ assuming
near-zero values; which
implies that transmission only occurs around $k_{t}\approx0$. We
show in Fig. \ref{fig:(a)-Reflection-and} the reflection and transmission
at 37 THz, by a TENZ material with $\epsilon_{t}=-0.001$ and $\epsilon_{z}=2.2$,
and by an IENZ material with $\epsilon_{r}=-0.001$, assuming in both
cases negligible losses. It is clear that the IENZ material exhibits
a very narrow transmission around $\theta_{i}\approx0^{\circ}$ only
due to evanescent waves (permittivity has a negative value) tunneling
through the subwavelength slab \cite{Salvoenz,LUOENZHM}, and the
transmission window dramatically diminishes as $\epsilon_{r}$ approaches
zero or $h$ increases, in accordance with the trend observed in \cite{alu2007epsilon}.
On the contrary, the TENZ slab exhibits large and stable transmission
over a wide range of incidence angles, inherently complying with the
flat wavevector dispersion relation in Eq. (\ref{blochtenz2}). Also,
one should point out that the TM$^{z}$ transmission in TENZ materials
occurs up to much larger incidence angles than TE$^{z}$ transmission,
which is identical to an IENZ slab's TE$^{z}$ transmission discussed
in \cite{alu2007epsilon}.
In principle the different properties illustrated in the preceding
simple example reveal the advantage of TENZ material over conventional
IENZ material in enhancing transmission under oblique TM$^{z}$ plane
wave incidence. For a more practical comparison, we report in Fig.
\ref{fig:(a)-Reflection-and2} the transmission and reflection for
two possible TENZ and IENZ materials at mid-infrared. We consider
a TENZ made of graphene-dielectric multilayer biased with $\mu_{c}=0.5$
eV, accounting for losses, and having total thickness of $h=Nd$ where
$d=50$ nm, at 37 THz. Under these conditions EMA estimates $\epsilon_{t}\approx-0.001-j0.031$
as seen from Fig. \ref{Fig__2}. The IENZ material is assumed to be
a heavily n-doped $\textrm{InAsSb}$ semiconductor \cite{ENZAdamsfunneling},
which is engineered via doping to exhibit low loss IENZ in this frequency
range, i.e., $\epsilon_{\textrm{InAsSb}}\approx-0.0001-j0.038$ at
$\approx$37 THz (experimentally shown in \cite{ENZAdamsfunneling}).
In graphene-based TENZ material we observe a stable transmission with
respect to the angle of incidence, and it is not affected much by
losses in graphene as deduced from the comparison of the lossy case
in Fig. \ref{fig:(a)-Reflection-and2} and the lossless case in Fig.
\ref{fig:(a)-Reflection-and}. The InAsSb thin slab, however, exhibits
a narrow angular range of transmission with higher sensitivity to
losses, i.e., as the imaginary part of $\epsilon_{r}$ is increased,
angular transmission is slightly broadened, especially as $h$ increases.
This indicates an advantage of using the graphene-based TENZ materials
in tuning and enhancing TM$^{z}$ plane wave transmission for wide
angles of incidence. On the other hand, losses in natural materials
or engineered metamaterials that exhibit IENZ behavior degrades the
performance considerably, and may require integration of gain materials
as in \cite{Salvoenz}.

\section{Conclusion}
{
We have reported a HM implementation at far- and mid-infrared frequencies
that comprises graphene-dielectric layers, and showed that EMA describes
the hyperbolic wavevector dispersion as well as the transition to
elliptic regime for specific conditions. Hyperbolic dispersion have
manifested mainly at far-infrared frequencies, where we have investigated
the propagating spectrum properties and discussed the effect of losses.
We also showed that hyperbolic and elliptic dispersion regimes are
associated to backward and forward wave propagation, respectively.
We have explored the tuning opportunities and design considerations
of the structure, as well as the translation from hyperbolic to elliptic
wavevector dispersion, and demonstrated a transitional state, TENZ,
at which the wavevector dispersion diagram becomes very flat. Furthermore,
we have demonstrated that a thin slab made by a TENZ material becomes
transparent to both TE$^{z}$ and TM$^{z}$ plane wave, with the interesting
characteristic that the transmission and reflection of TM$^{z}$ waves
are stable with respect to the incident angle, in contrast to what
happens in conventional IENZ materials. This property can be utilized
in designing ultra-thin films for tunable infrared applications.}

\end{document}